\documentclass[final,5p,times,twocolumn,sort&compress]{elsarticle}

\usepackage{graphicx}

\usepackage{amsmath}
\usepackage{amssymb}

\journal{Physics Letters B}

\begin{document}

\begin{frontmatter}

\title{Topological black holes in ${\mathfrak {su}}(N)$ Einstein-Yang-Mills theory \\ with a negative cosmological constant}

\author[jeb]{J. Erik Baxter}
\ead{E.Baxter@shu.ac.uk}

\author[ew]{Elizabeth Winstanley}
\ead{E.Winstanley@sheffield.ac.uk}

\address[jeb]{Norfolk Building, Sheffield Hallam University, 1 Howard Street, Sheffield. S1 1WB United Kingdom}
\address[ew]{Consortium for Fundamental Physics, School of Mathematics and Statistics,
The University of Sheffield, \\ Hicks Building, Hounsfield Road, Sheffield. S3 7RH United Kingdom}

\date{\today}

\begin{abstract}
We investigate the phase space of topological black hole solutions of ${\mathfrak {su}}(N)$ Einstein-Yang-Mills theory in anti-de Sitter space with a purely magnetic gauge potential.
The gauge field is described by $N-1$ magnetic gauge field functions $\omega _{j}$, $j=1,\ldots , N-1$.
For ${\mathfrak {su}}(2)$ gauge group, the function $\omega _{1}$ has no zeros. This is no longer the case when we consider a larger gauge group.
The phase space of topological black holes is considerably simpler than for the corresponding spherically symmetric black holes, but for $N>2$ and a flat event horizon, there exist solutions where at least one of the $\omega _{j}$ functions has one or more zeros.
For most of the solutions, all the $\omega _{j}$ functions have no zeros, and at least some of these are linearly stable.
\end{abstract}


\end{frontmatter}

\section{Introduction}
\label{sec:intro}

Many properties of black holes in asymptotically anti-de Sitter (adS) space-time are rather different from those of black holes in asymptotically flat space-time.
For example, stationary vacuum black holes in four-dimensional asymptotically flat space-time must have spherical event horizon topology \cite{Hawking:1971vc, Hawking:1973uf}
but in four-dimensional adS space-time, vacuum black holes with other event horizon topologies exist (see, for example,
\cite{Birmingham:1998nr, Brill:1997mf, Lemos:1994fn, Lemos:1994xp, Lemos:1995cm, Vanzo:1997gw, Cai:1996eg, Mann:1996gj, Smith:1997wx, Mann:1997zn}).

Many properties of black holes with nontrivial matter field hair in asymptotically adS space-time also differ from the properties of their asymptotically flat counterparts.
In ${\mathfrak {su}}(2)$ Einstein-Yang-Mills (EYM) theory, spherically symmetric, asymptotically flat ``coloured'' black holes \cite{Bizon:1990sr} are unstable under linear, spherically symmetric, perturbations \cite{Straumann:1990as}.
In contrast, there exist stable, spherically symmetric, asymptotically adS, black hole solutions of ${\mathfrak {su}}(2)$ EYM with a negative cosmological constant \cite{Winstanley:1998sn, Bjoraker:1999yd, Bjoraker:2000qd}.
Analogues of these spherically symmetric, asymptotically adS black holes with nonspherical event horizon topology also exist when the gauge group is ${\mathfrak {su}}(2)$ \cite{VanderBij:2001ia}. Furthermore, these topological EYM black holes are linearly stable \cite{VanderBij:2001ia}.

Since the discovery of stable black holes in ${\mathfrak {su}}(2)$ EYM in adS, black hole solutions with the larger ${\mathfrak {su}}(N)$ gauge group have been extensively studied (see \cite{Winstanley:2008ac, Winstanley:2015loa} for reviews).
For all $N$, there exist stable, spherically symmetric, purely magnetic, asymptotically adS black holes \cite{Baxter:2007at, Baxter:2008pi, Baxter:2015gfa}.
For ${\mathfrak {su}}(N)$ gauge group, the purely magnetic gauge field is described by $N-1$ independent functions, so these stable black holes have
an unlimited amount of gauge field ``hair''.

A natural question is whether there exist analogues of the purely magnetic, spherically symmetric black holes in ${\mathfrak {su}}(N)$ EYM in adS with nonspherical event horizon topology.
Recently this question was answered in the affirmative \cite{Baxter:2014nka}, in two regimes: (a) for a negative cosmological constant with sufficiently large magnitude and (b) in a neighbourhood of an embedded ${\mathfrak {su}}(2)$ solution.
Furthermore, the proof of the existence of stable spherically symmetric ${\mathfrak {su}}(N)$ EYM black holes in adS \cite{Baxter:2015gfa} can be extended to these topological ${\mathfrak {su}}(N)$ EYM black holes, showing that at least some purely magnetic topological black holes, in the intersection of the regimes (a) and (b) above, are linearly stable \cite{Baxter:2015xfa}.

The analytic work in \cite{Baxter:2014nka,Baxter:2015xfa} proves the {\em {existence}} of stable topological black holes in ${\mathfrak {su}}(N)$ EYM in adS, but does not tell us about the nature of the phase space of topological black hole solutions.
In the spherically symmetric case, the phase space of purely magnetic black hole solutions of ${\mathfrak {su}}(N)$ EYM in adS has a very rich structure \cite{Baxter:2007au}.
In this letter we explore the phase space of numerical topological black hole solutions of ${\mathfrak {su}}(N)$ EYM in adS, drawing comparisons with the spherically symmetric solution space studied in \cite{Baxter:2007au}.  We begin in section \ref{sec:EYM} with a review of the salient features of ${\mathfrak {su}}(N)$ EYM in adS before presenting our numerical results in section \ref{sec:BH} and our conclusions in section \ref{sec:conc}.

\section{${\mathfrak {su}}(N)$ Einstein-Yang-Mills theory in adS}
\label{sec:EYM}

We consider Einstein-Yang-Mills (EYM) theory in four space-time dimensions, having the action:
\pagebreak
\begin{equation}
S= \frac {1}{2} \int d^{4}x \, {\sqrt {-g}} \left[  R-2\Lambda -
{\mbox {Tr}} \, F_{\alpha \beta }F^{\alpha \beta } \right] ,
\label{eq:EYMaction}
\end{equation}
where $R$ is the Ricci scalar and $\Lambda <0$ the cosmological constant.
Here and throughout this paper we use units in which $8\pi G = c=1$.
We also set the gauge coupling constant equal to unity.
The nonabelian gauge field strength $F_{\mu \nu }$ is given in terms of the gauge field potential $A_{\mu }$ as follows:
\begin{equation}
F_{\mu \nu } = \partial _{\mu } A_{\nu } - \partial _{\nu } A_{\mu } +  \left[ A_{\mu },A_{\nu } \right] .
\end{equation}
Varying the action (\ref{eq:EYMaction}) gives the EYM field equations:
\begin{eqnarray}
R_{\alpha \beta } - \frac {1}{2} R g_{\alpha \beta } + \Lambda g_{\alpha \beta } & = &  T_{\alpha \beta } ,
\nonumber
\\
\nabla _{\alpha } F^{\alpha }{}_{\beta } +
\left[ A_{\alpha }, F{}^{\alpha }{}_{\beta } \right] & = & 0 ,
\label{eq:feqns}
\end{eqnarray}
where the stress-energy tensor of the Yang-Mills field is:
\begin{equation}
T_{\alpha \beta } = \, {\mbox {Tr}} \, F_{\alpha \lambda } F^{\lambda }{}_{\beta } - \frac {1}{4} g_{\alpha \beta }
{\mbox {Tr}} \, F_{\lambda \sigma} F^{\lambda \sigma } ,
\end{equation}
with ${\mbox {Tr}}$ denoting a Lie algebra trace.

In this letter we are interested in asymptotically anti-de Sitter (adS) topological black holes.
The metric ansatz we employ takes the form:
\begin{equation}
ds^{2} = - \mu (r) S (r)^{2} \, dt^{2} + \left[ \mu (r) \right] ^{-1} dr^{2}
+ r^{2} \left( d\theta ^{2} + f_{k}^{2}(\theta ) \, d\phi ^{2} \right) ,
\label{eq:topmetric}
\end{equation}
where the metric functions $\mu (r)$ and $S(r)$ depend on the radial co-ordinate $r$ only.
The metric function $\mu (r)$ can be written in an alternative form in terms of a new function $m(r)$:
\begin{equation}
\mu (r) = k - \frac {2m(r)}{r} - \frac {\Lambda r^{2}}{3} .
\label{eq:mu}
\end{equation}
In (\ref{eq:topmetric}, \ref{eq:mu}), the constant $k$ can take the values $\{ 1, 0, -1\}$.
The form of the function $f_{k}$ depends on $k$ as follows:
\begin{equation}
f_{k}(\theta ) =
\left\{
\begin{array}{ll}
\sin \theta, \qquad & k=1,
\\
\theta, \qquad & k=0,
\\
\sinh \theta, \qquad & k=-1.
\end{array}
\right.
\label{eq:f}
\end{equation}
When $k=1$, the metric (\ref{eq:topmetric}) is spherically symmetric, with the $t={\text {constant}}$, $r={\text {constant}}$ hypersurfaces being two-spheres. For $k=0$, the $t={\text {constant}}$, $r={\text {constant}}$ hypersurfaces are two-dimensional Euclidean spaces, while for $k=-1$ these hypersurfaces have constant negative curvature (see
\cite{Birmingham:1998nr, Brill:1997mf, Lemos:1994fn, Lemos:1994xp, Lemos:1995cm, Vanzo:1997gw, Cai:1996eg, Mann:1996gj, Smith:1997wx, Mann:1997zn}
for further details).

An appropriate ansatz for a purely magnetic ${\mathfrak {su}}(N)$ Yang-Mills gauge field potential on the space-time with metric (\ref{eq:topmetric}) is
\cite{VanderBij:2001ia, Baxter:2014nka, Kunzle:1991}
\begin{eqnarray}
A_{\alpha } \, d x^{\alpha } & = & \frac {1}{2} \left( C - C^{H} \right) d\theta
\nonumber \\ & &
- \frac {i}{2} \left[ \left( C + C^{H} \right) f_{k}(\theta ) + D \frac {d f_{k}(\theta )}{d\theta } \right] d\phi ,
\label{eq:gaugepot}
\end{eqnarray}
where $C$ and $D$ are $N\times N$ matrices.
The matrix $C$ has hermitian conjugate $C^{H}$ and is upper-triangular, with nonzero entries only immediately above the diagonal.
These nonzero entries can be written in terms of $N-1$ functions of $r$ only:
\begin{equation}
C_{j,j+1} = \omega _{j}(r), \qquad j=1,\ldots , N-1 .
\end{equation}
The matrix $D$ is constant and diagonal:
\begin{equation}
D = {\mbox {Diag}} \{ N-1, N -3, \ldots, -N +3 , -N +1 \} .
\end{equation}
The ansatz (\ref{eq:gaugepot}) reduces in the case $k=1$ to that in \cite{Kunzle:1991} for a purely magnetic spherically-symmetric ${\mathfrak {su}}(N)$ gauge field potential. For ${\mathfrak {su}}(2)$ EYM, it is the ansatz used in \cite{VanderBij:2001ia} for topological black holes. The general form (\ref{eq:gaugepot})
was derived in \cite{Baxter:2014nka}.
The ${\mathfrak {su}}(N)$ gauge potential (\ref{eq:gaugepot}) is therefore described by the $N-1$ magnetic field functions $\omega _{j}(r)$, $j=1, \ldots , N-1$.

With the form of the gauge potential (\ref{eq:gaugepot}) and the metric ansatz (\ref{eq:topmetric}, \ref{eq:mu}), the field equations (\ref{eq:feqns}) simplify to the following Einstein equations \cite{Baxter:2014nka}:
\begin{equation}
m'  =  \mu G + r^{2} p_{k}, \qquad
\frac {S'}{S}  =  \frac {2G}{r},
\label{eq:EE}
\end{equation}
with
\begin{equation}
G  =  \sum _{j=1}^{N-1} \omega _{j}'^{2}, \qquad
p_{k}  = \frac {1}{4r^{4}}  \sum _{j=1}^{N} \left[ \omega _{j}^{2} - \omega _{j-1}^{2} - k \left( N +1 -2j \right) \right] ^{2},
\end{equation}
and $N-1$ coupled Yang-Mills equations
\begin{equation}
0 = r^{2} \mu \omega _{j}'' +
\left( 2m - 2r^{3} p_{k} - \frac {2\Lambda r^{3}}{3} \right) \omega _{j}'
+ W_{k,j} \omega _{j},
\label{eq:YME}
\end{equation}
where
\begin{equation}
W_{k,j} = k - \omega _{j}^{2} + \frac {1}{2} \left( \omega _{j-1}^{2} + \omega _{j+1}^{2} \right) .
\end{equation}
As in the $k=1$ case, the equation (\ref{eq:EE}) for the metric variable $S(r)$ decouples from the remaining equations.
Furthermore, the field equations (\ref{eq:EE}, \ref{eq:YME}) are invariant under the transformation $\omega _{j}\rightarrow - \omega _{j}$ for each $j$ independently, and also under $j\rightarrow N-j$ for all $j$.

We are interested in topological black holes with a regular event horizon at $r=r_{h}$, where $\mu (r_{h})=0$ and $\mu '(r_{h})>0$.
We assume that the field variables $m(r)$, $S(r)$ and $\omega _{j}(r)$ have regular Taylor series expansions in a neighbourhood of $r=r_{h}$, of the form
\begin{eqnarray}
m(r) & = &
m(r_{h}) + m'(r_{h}) \left( r-r_{h} \right) + O\left( r-r_{h} \right) ^{2},
\nonumber \\
S(r) & = & S(r_{h}) + S'(r_{h}) \left( r-r_{h} \right) + O\left( r-r_{h} \right) ^{2},
\nonumber \\
\omega _{j}(r) & = & \omega _{j}(r_{h}) + \omega _{j}'(r_{h})\left( r-r_{h} \right) + O\left( r-r_{h} \right) ^{2}.
\label{eq:horizon}
\end{eqnarray}
The condition $\mu (r_{h})=0$ fixes $m(r_{h})$ to be
\begin{equation}
m(r_{h}) = \frac {r_{h}}{2} \left( k - \frac {\Lambda r_{h}^{2}}{3} \right) .
\end{equation}
The parameters $S(r_{h})$ and $\omega _{j}(r_{h})$ are {\it {a priori}} arbitrary; the value of $S(r_{h})$ will be fixed by the boundary conditions at infinity.
From the field equations (\ref{eq:EE}, \ref{eq:YME}) the values of $m'(r_{h})$, $\omega _{j}'(r_{h})$ and $S'(r_{h})$ are fixed in terms of $r_{h}$, $\Lambda $ and the values of the magnetic gauge field functions on the horizon $\omega _{j}(r_{h})$ \cite{Baxter:2014nka}.
For a regular event horizon at $r=r_{h}$, we require $\mu '(r_{h})>0$ and this implies that
\begin{equation}
m'(r_{h}) = r_{h}^{2}p_{k}(r_{h}) < \frac {1}{2} \left( k - \Lambda r_{h}^{2} \right) ,
\label{eq:EHcond}
\end{equation}
which places a (weak) constraint on the values of the magnetic gauge field functions on the horizon $\omega _{j}(r_{h})$.
For the $k=-1$ case, the relation (\ref{eq:EHcond}) implies the existence of a minimum value of $\left| \Lambda \right| $ for fixed $r_{h}$:
\begin{equation}
\left| \Lambda \right| > \frac {1}{r_{h}^{2}} \left( 1 + \frac {N(N-1)(N+1)}{6r_{h}^{2}} \right) .
\label{eq:rhbound}
\end{equation}

As $r\rightarrow \infty $, the field variables have the following expansions \cite{Baxter:2014nka}:
\begin{eqnarray}
m(r)  & =  &
M + O(r^{-1}), \qquad
S(r) = 1 + O(r^{-1}), \nonumber \\
\omega _{j}(r) & =  & \omega _{j,\infty } + O(r^{-1}),
\label{eq:infinity}
\end{eqnarray}
where $M$ and $\omega _{j, \infty }$ are arbitrary constants.
The value of the metric function $S(r_{h})$ on the horizon is fixed by the requirement that $S(r)\rightarrow 1 $ as $r\rightarrow \infty $.
The local existence of solutions of the field equations (\ref{eq:EE}, \ref{eq:YME}) satisfying the boundary conditions (\ref{eq:horizon}, \ref{eq:infinity})
is proven in \cite{Baxter:2014nka}.

The field equations (\ref{eq:EE}, \ref{eq:YME}) possess a trivial solution when all the magnetic gauge field functions are set to vanish identically
$\omega _{j}(r) \equiv 0$.
In this case $S(r)\equiv 1$ and the other metric function takes the form
\begin{equation}
\mu (r) = k - \frac {2M}{r} + \frac {Q^{2}}{r^{2}} - \frac {\Lambda r^{2}}{3},
\end{equation}
where $M$ is an arbitrary constant and the magnetic charge $Q$ is given by
\begin{equation}
Q^{2} = \frac {k^{2}}{6} N \left( N +1 \right) \left( N -1 \right) .
\label{eq:charge}
\end{equation}
For $k=0$, we have $Q=0$ and the solution is the embedded Schwarzschild-adS black hole with planar event horizon topology.
For $k=-1$, the magnetic charge $Q>0$ and we have an embedded (magnetically-charged) Reissner-Nordstr\"om-adS black hole with constant negative curvature horizon.

\section{Topological black hole solutions}
\label{sec:BH}

We now integrate the field equations (\ref{eq:EE}, \ref{eq:YME}) numerically. We begin our integration close to the event horizon $r=r_{h}$, using the expansion of the field variables (\ref{eq:horizon}) as initial conditions. We then integrate for increasing $r$ until either the solution approaches the boundary conditions (\ref{eq:infinity}) to within a suitable tolerance or the solution becomes singular.
The field equations (\ref{eq:EE}, \ref{eq:YME}) are invariant under the transformation $\omega _{j}(r) \rightarrow - \omega _{j}(r)$ for each $j$ independently, so without loss of generality we may restrict our attention to values of the parameters such that $\omega _{j}(r_{h})>0$
for all $j$.

In this section we plot the phase spaces of topological black hole solutions of the field equations, fixing $r_{h}=1$, so that all length scales are in units of the event horizon radius, and considering various $\Lambda $.
The plots show the spaces of initial parameters $v_{j}=\omega _{j}(r_{h})$ which determine the solutions, and we explore the values of $v_{j}$ for which there is a regular event horizon, so that (\ref{eq:EHcond}) holds.
In each plot,  we show the region of the parameter space for which there are nontrivial topological black hole solutions and label these solutions by the quantities $n_{j}$, with $n_{j}$ being the number of zeros of the magnetic gauge field function $\omega _{j}$.
We review the numerical solutions for ${\mathfrak {su}}(2)$ gauge group \cite{VanderBij:2001ia} before discussing our new solutions for ${\mathfrak {su}}(3)$ gauge group.

\subsection{${\mathfrak {su}}(2)$ topological black holes}
\label{sec:su2}

\begin{figure*}[t]
\begin{center}
\begin{tabular}{cc}
\includegraphics[width=0.95\columnwidth]{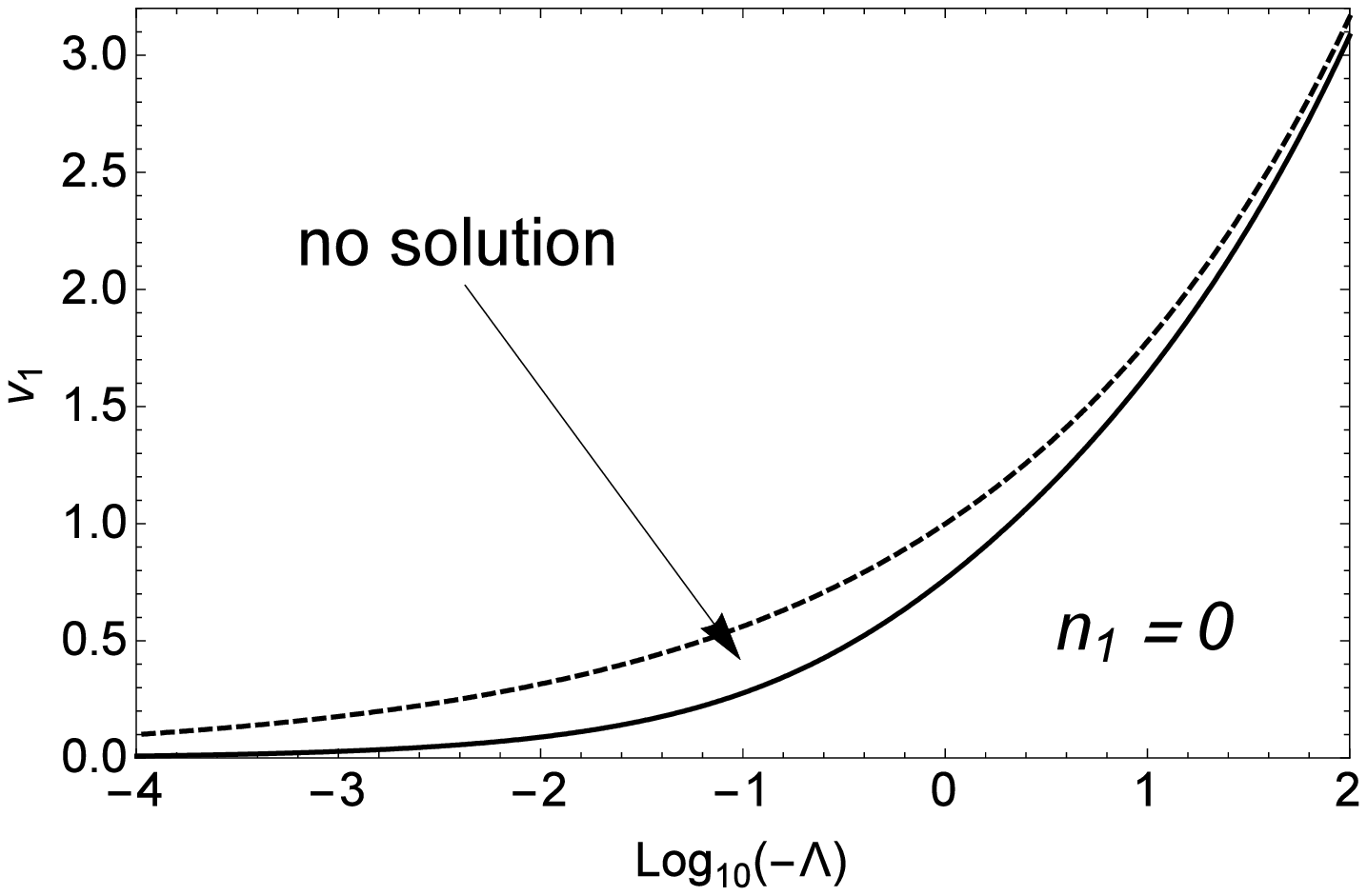} &
\includegraphics[width=0.95\columnwidth]{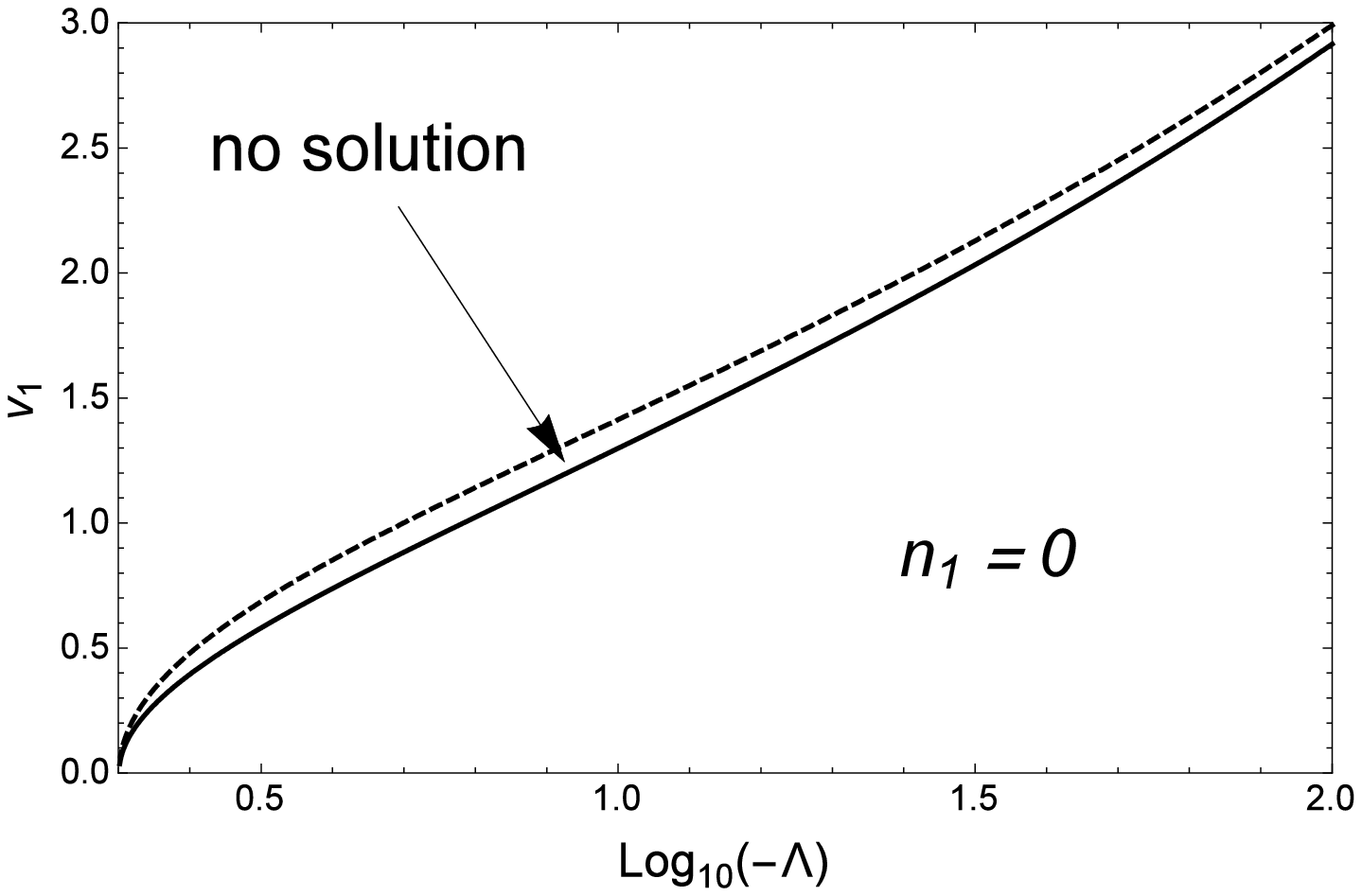}
\\
(a) $k=0$
&
(b) $k=-1$
\end{tabular}
\end{center}
\caption{Phase space of topological black hole solutions of ${\mathfrak {su}}(2)$ EYM for  (a) $k=0$, (b) $k=-1$.  In both plots we have fixed the event horizon
radius to be $r_{h}=1$ and varied $\left| \Lambda \right|$. Black hole solutions exist in the regions below the solid curves. For parameter values above the solid curves there are no solutions.  The region bounded by dashed and solid curves denotes that region of parameter space where the condition (\ref{eq:EHcond}) for a regular event horizon holds, but we did not find a suitable numerical solution.   All topological black hole solutions are such that $\omega _{1}(r)$ has no zeros so that $n_{1}=0$.}
\label{fig:su2}
\end{figure*}

Topological black hole solutions of the ${\mathfrak {su}}(2)$ EYM field equations (\ref{eq:EE}, \ref{eq:YME}) were first found in \cite{VanderBij:2001ia}.
In this case the YM field is described by a single gauge field function $\omega _{1}(r)$.  In figure~\ref{fig:su2} we plot the phase space of solutions for $k=0$ and $k=-1$ in plots (a) and (b) respectively, with $r_{h}=1$ and varying cosmological constant $\Lambda $.
The black holes are parameterized by $\Lambda $ and $v_{1}=\omega _{1}(r_{h})$, the value of the single magnetic gauge field function on the horizon.
In figure 1 we show the space of the parameters $(\Lambda ,v_{1})$ satisfying the condition (\ref{eq:EHcond}) for a regular black hole event horizon at $r=r_{h}$. This is the region underneath the dotted curves.
For some values of the parameters $(\Lambda , v_{1})$ such that (\ref{eq:EHcond}) is satisfied, the numerical solution of the field equations (\ref{eq:EE}, \ref{eq:YME}) becomes singular at some $r<\infty$.  Such points are contained in the region labelled ``no solution'', bounded by the dotted and solid curves.

Nontrivial topological black holes exist in the regions below the solid curves in the plots in figure 1.
The line $v_{1}=0$ corresponds to the trivial embedded Reissner-Nordstr\"om-adS black hole with magnetic charge given by (\ref{eq:charge}).
As shown in \cite{VanderBij:2001ia}, for all the solutions for both $k=0$ and $k=-1$ the single magnetic gauge field function $\omega _{1}(r)$ has no zeros, so $n_{1}=0$.

For $k=0$, we find nontrivial solutions for all values of $\Lambda $ with $r_{h}=1$, but for $\left| \Lambda \right| $ smaller than $10^{-3}$ the range of values of $v_{1}$ for which there are solutions is very small, as in this case the condition (\ref{eq:EHcond}) reduces to
$v_{1}^{4} < -\Lambda r_{h}^{4}$.
In accordance with this inequality, as $\left| \Lambda \right| $ increases, we find an increasingly large range of $v_{1}$ for which nontrivial black hole solutions exist.

For $k=-1$ and $r_{h}=1$, the minimum value of $\left| \Lambda \right| $ for which there is a regular event horizon, given by (\ref{eq:rhbound}), reduces to
$\left| \Lambda \right| > 2$. As in the $k=0$ case, as $\left| \Lambda \right| $ increases for $k=-1$, we find an increasing range of values of $v_{1}$ for which there are nontrivial black hole solutions.

The phase spaces in figure~\ref{fig:su2} for $k=0$ and $k=-1$ are much simpler than the corresponding phase space of spherically symmetric ($k=1$) black holes (see, for example, \cite{Baxter:2007au}).
In the $k=1$ case, for fixed $r_{h}$ and sufficiently large $\left| \Lambda \right| $, all the black holes are such that $\omega _{1}(r)$ has no zeros.
However, as $\left| \Lambda \right| $ decreases, $n_{1}$ increases (for example, with $\left| \Lambda \right| = 10^{-4}$, there are $k=1$ black holes with $\omega _{1}(r)$ having up to four zeros \cite{Baxter:2007au}).  As $\left| \Lambda \right| \rightarrow 0$, the phase space of $k=1$ black holes fragments and reduces to the asymptotically flat ``coloured'' black holes \cite{Bizon:1990sr}.
If $k\neq 1$, there is no asymptotically flat limit.  For $k=-1$ the phase space of topological black holes simply ends when $\left| \Lambda \right| <2$, while for $k=0$ the phase space shrinks to zero size as $\left| \Lambda \right| \rightarrow 0$.

In \cite{VanderBij:2001ia} it is shown that all the $k=0$ topological black hole solutions shown in figure \ref{fig:su2} are stable under linear, spherically symmetric perturbations.  For $k=-1$, solutions for which $\omega _{1,\infty }>1$ in (\ref{eq:infinity}) are shown to be stable \cite{VanderBij:2001ia}.

\subsection{${\mathfrak {su}}(3)$ topological black holes}
\label{sec:su3}

All the ${\mathfrak {su}}(2)$ solutions shown in figure \ref{fig:su2} can be embedded as solutions of ${\mathfrak {su}}(N)$ EYM theory for any $N$ \cite{Baxter:2014nka}.  Setting
\begin{equation}
\omega _{j}(r) = {\sqrt {j\left( N-j\right) }} \omega (r),
\label{eq:su2embedded}
\end{equation}
and defining rescaled quantities ${\tilde {r}}$, ${\tilde {m}}$, ${\tilde {\Lambda }}$ as follows:
\begin{equation}
{\tilde {r}} = {\tilde {N}}^{-\frac {1}{2}} r, \qquad
{\tilde {m}} = {\tilde {N}}^{-\frac {1}{2}} m, \qquad
{\tilde {\Lambda }} = {\tilde {N}} \Lambda ,
\end{equation}
where
\begin{equation}
{\tilde {N}} = \frac {1}{6} N \left( N - 1 \right) \left( N + 1 \right) ,
\end{equation}
it is found that $\omega $ and ${\tilde {m}}$, considered as functions of ${\tilde {r}}$, satisfy the ${\mathfrak {su}}(2)$ EYM field equations (\ref{eq:EE}, \ref{eq:YME}) with cosmological constant ${\tilde {\Lambda }}$ \cite{Baxter:2014nka}.
In \cite{Baxter:2014nka} it is proven that there exist nontrivial (that is, nonembedded) ${\mathfrak {su}}(N)$ topological black holes in a neighbourhood of these embedded ${\mathfrak {su}}(2)$ black holes.

\begin{figure*}[t]
\begin{center}
\begin{tabular}{cc}
\includegraphics[width=0.95\columnwidth]{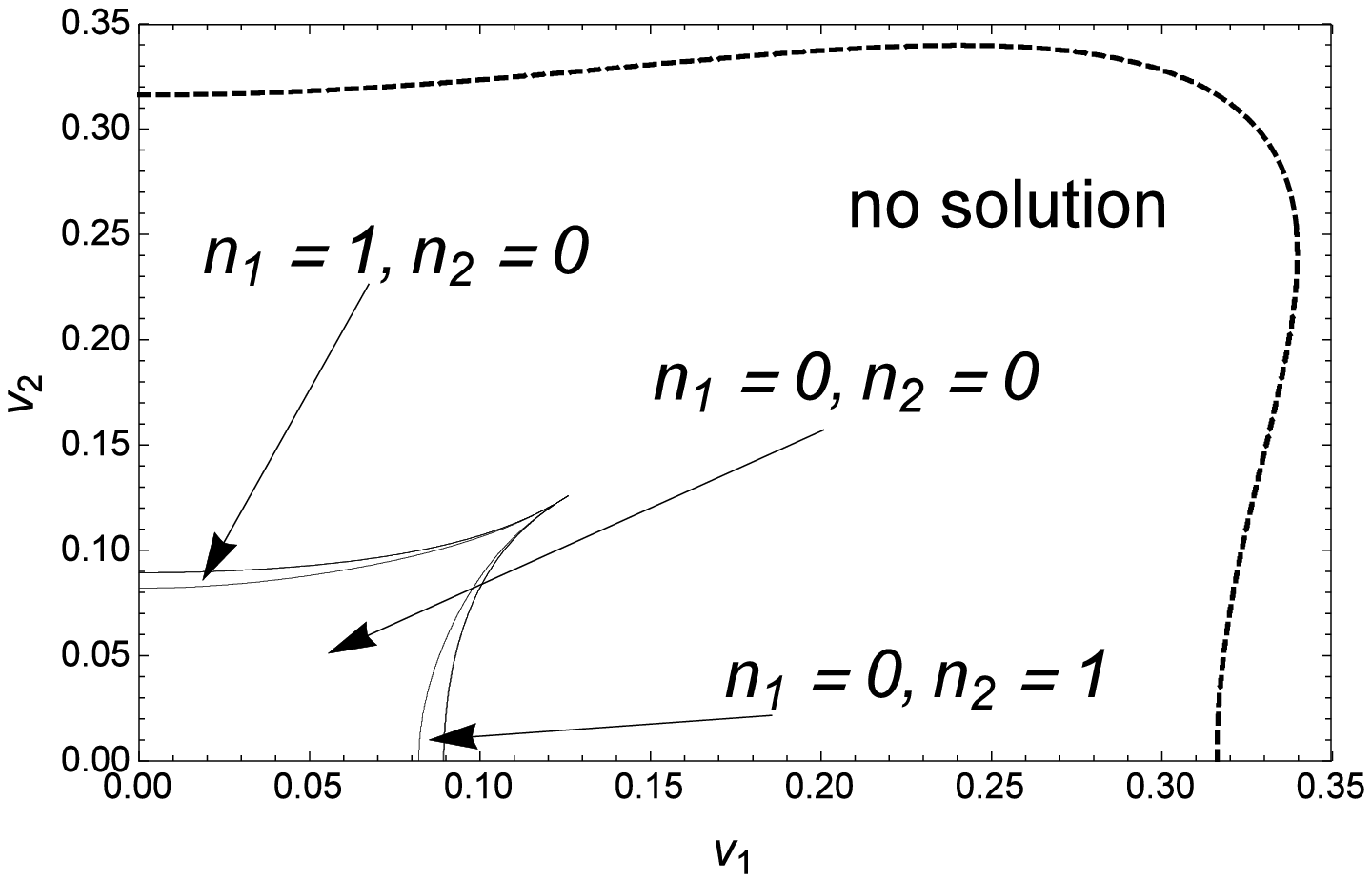}  &
\includegraphics[width=0.95\columnwidth]{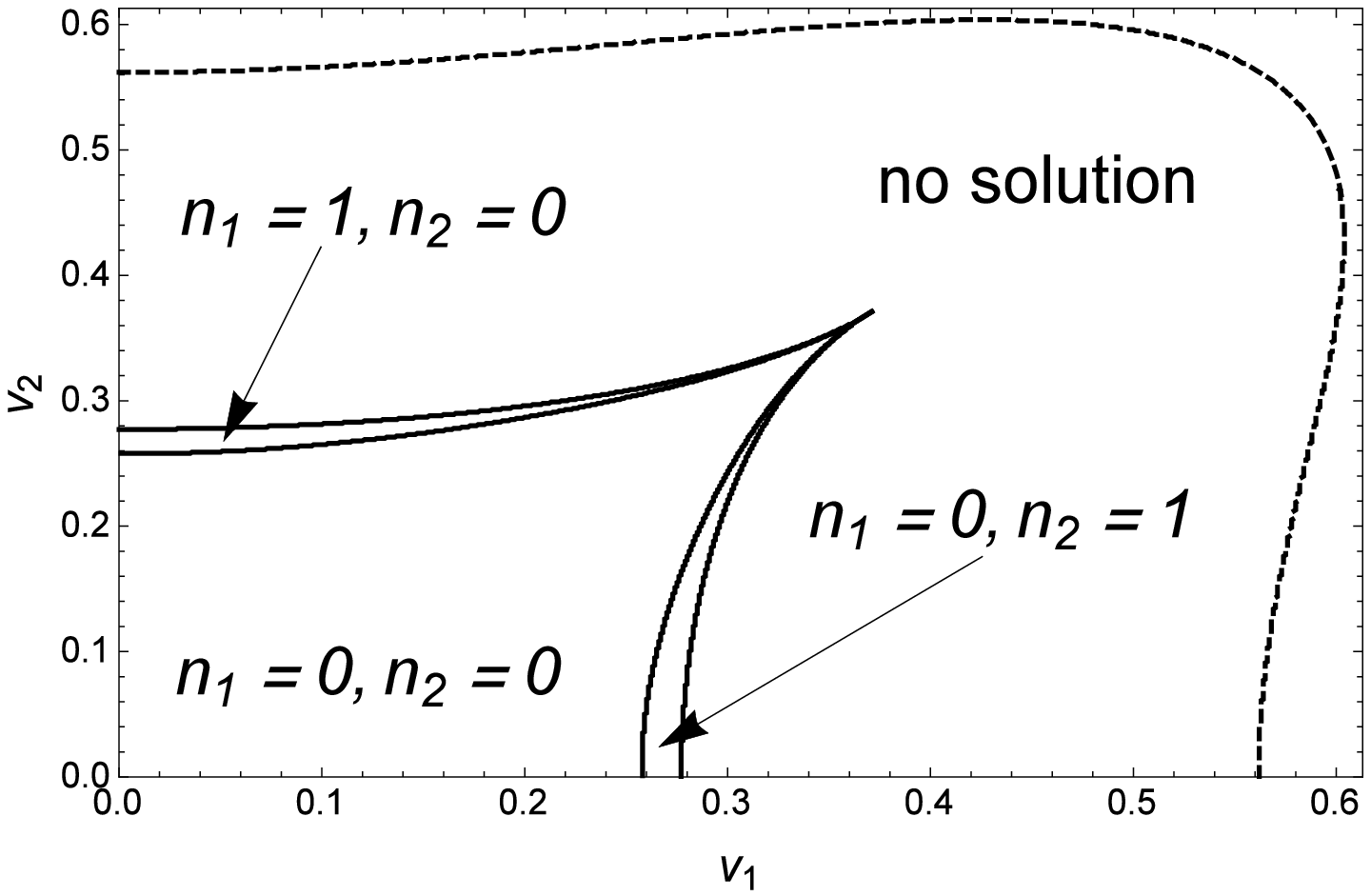}  \\
(a) $\Lambda = -0.01$ & (b) $\Lambda = -0.1$
\\
\includegraphics[width=0.95\columnwidth]{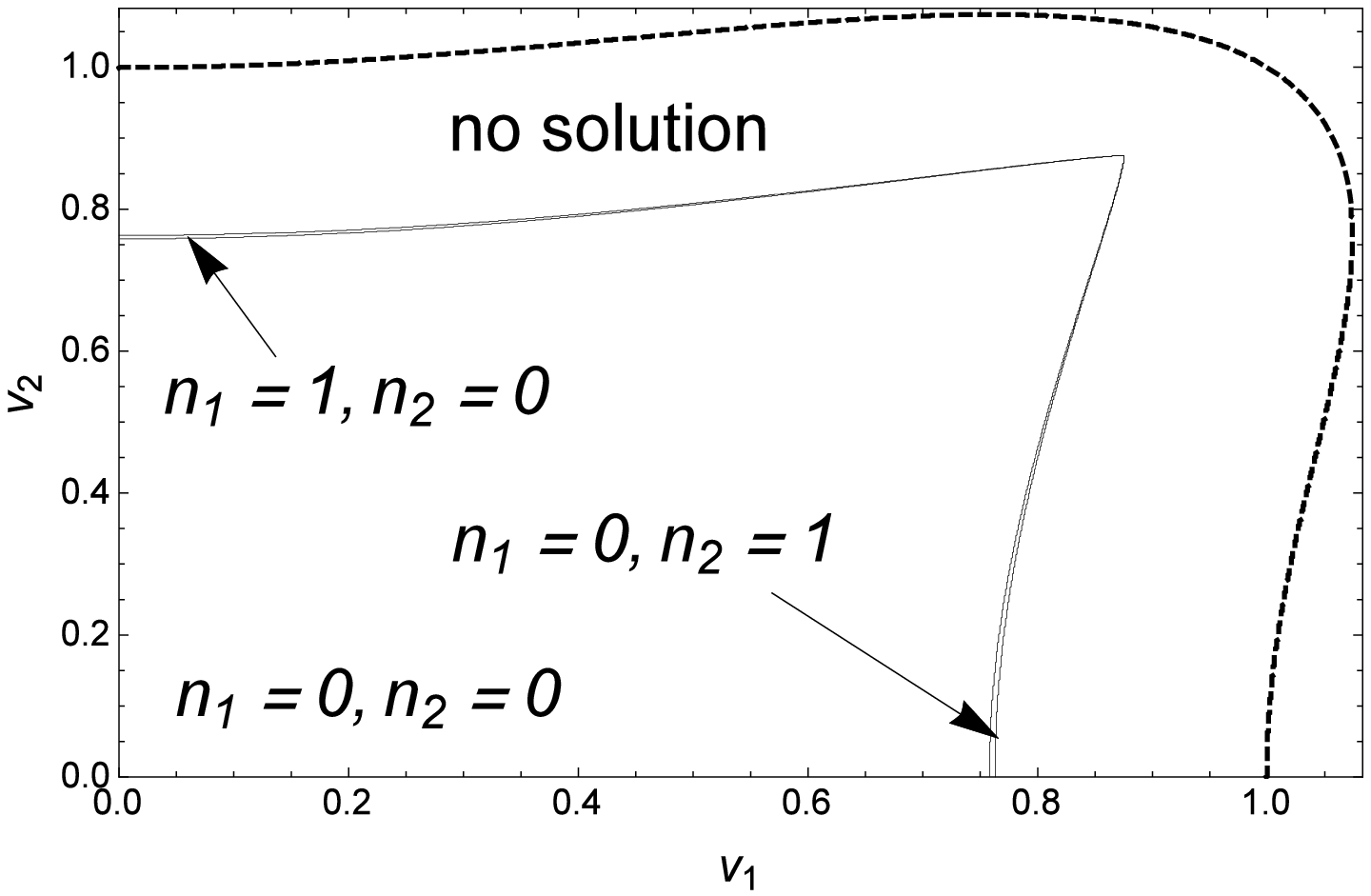}  &
\includegraphics[width=0.95\columnwidth]{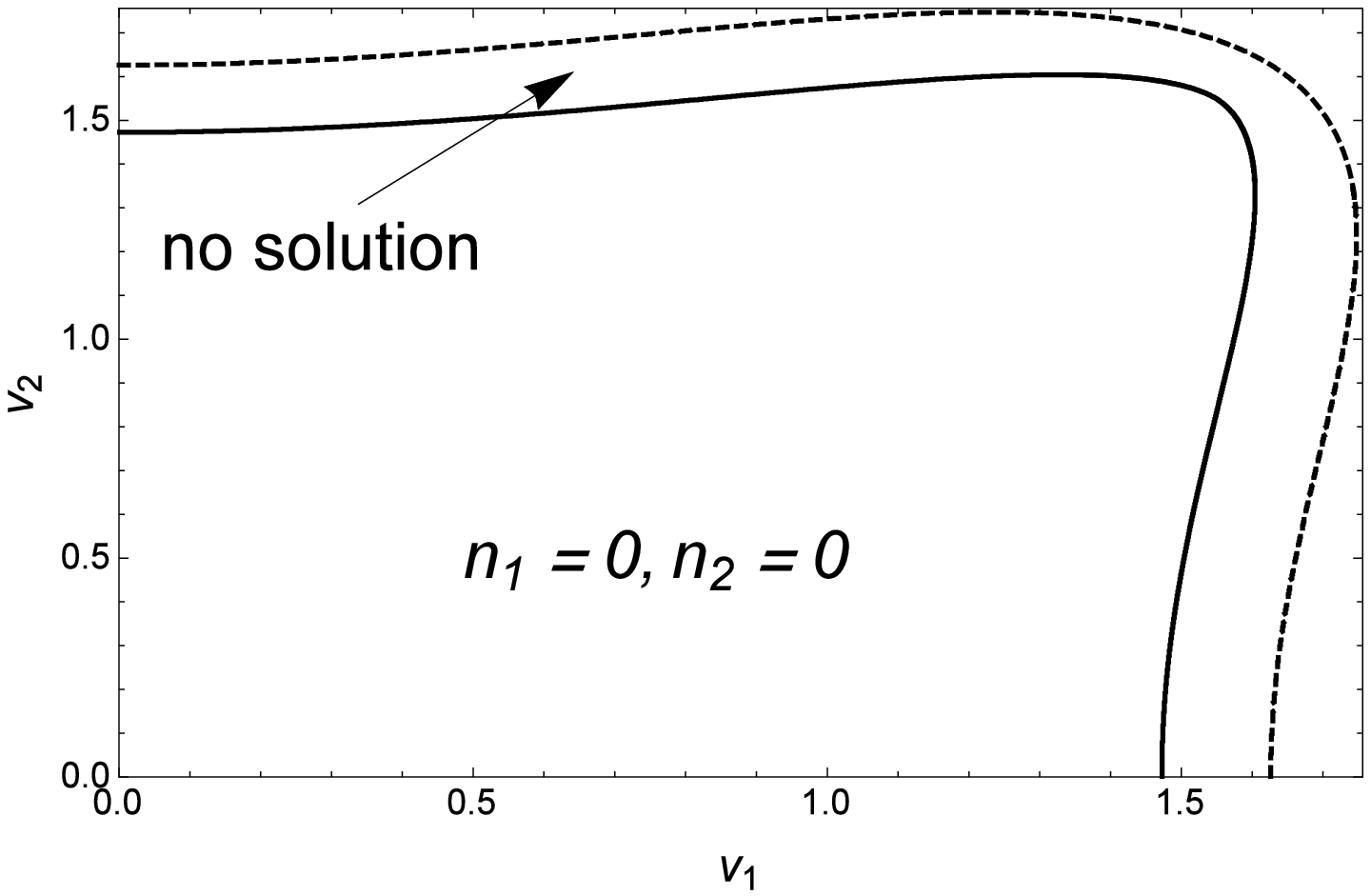}   \\
(c) $\Lambda = -1$ & (d) $\Lambda = -7$
\end{tabular}
\end{center}
\caption{Phase space of topological black hole solutions of ${\mathfrak {su}}(3)$ EYM for $k=0$. We have fixed the event horizon
radius to be $r_{h}=1$ and  $ \Lambda $ to have the values (a)~$-0.01$, (b)~$-0.1$, (c)~$-1$ and (d)~$-7$. Black hole solutions exist in the region bounded by the axes and the solid curves. For parameter values outside the solid curves there are no solutions.  The region bounded by dashed and solid curves denotes that region of parameter space where the condition (\ref{eq:EHcond}) for a regular event horizon holds, but we did not find a suitable numerical solution.
Regions where there are nontrivial topological black hole solutions are labelled according to the number of zeros $n_{1}$, $n_{2}$ of the magnetic gauge field functions $\omega _{1}$, $\omega _{2}$ respectively.}
\label{fig:su3k=0}
\end{figure*}

\begin{figure*}[t]
\begin{center}
\begin{tabular}{cc}
\includegraphics[width=0.95\columnwidth]{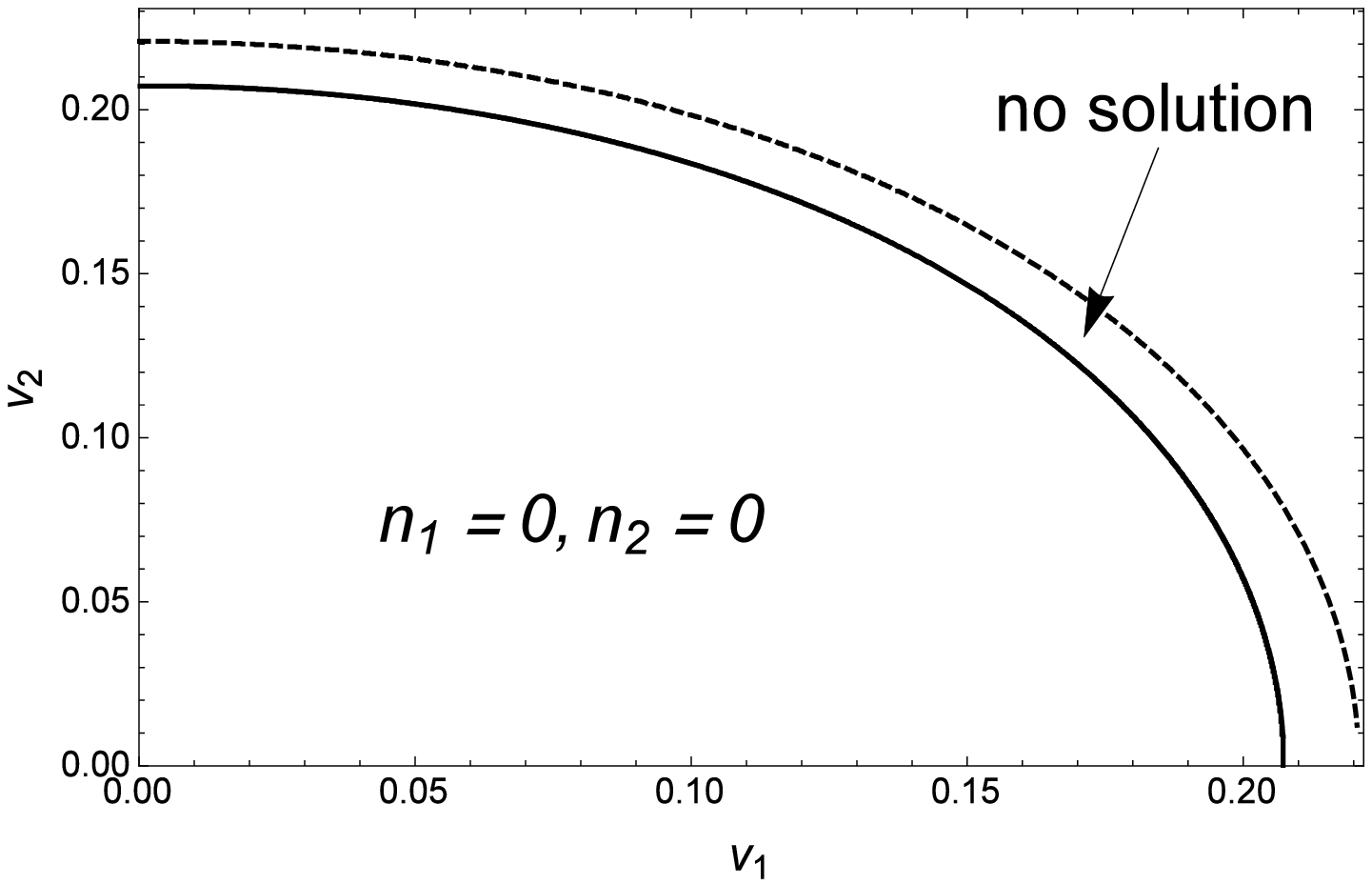}  &
\includegraphics[width=0.95\columnwidth]{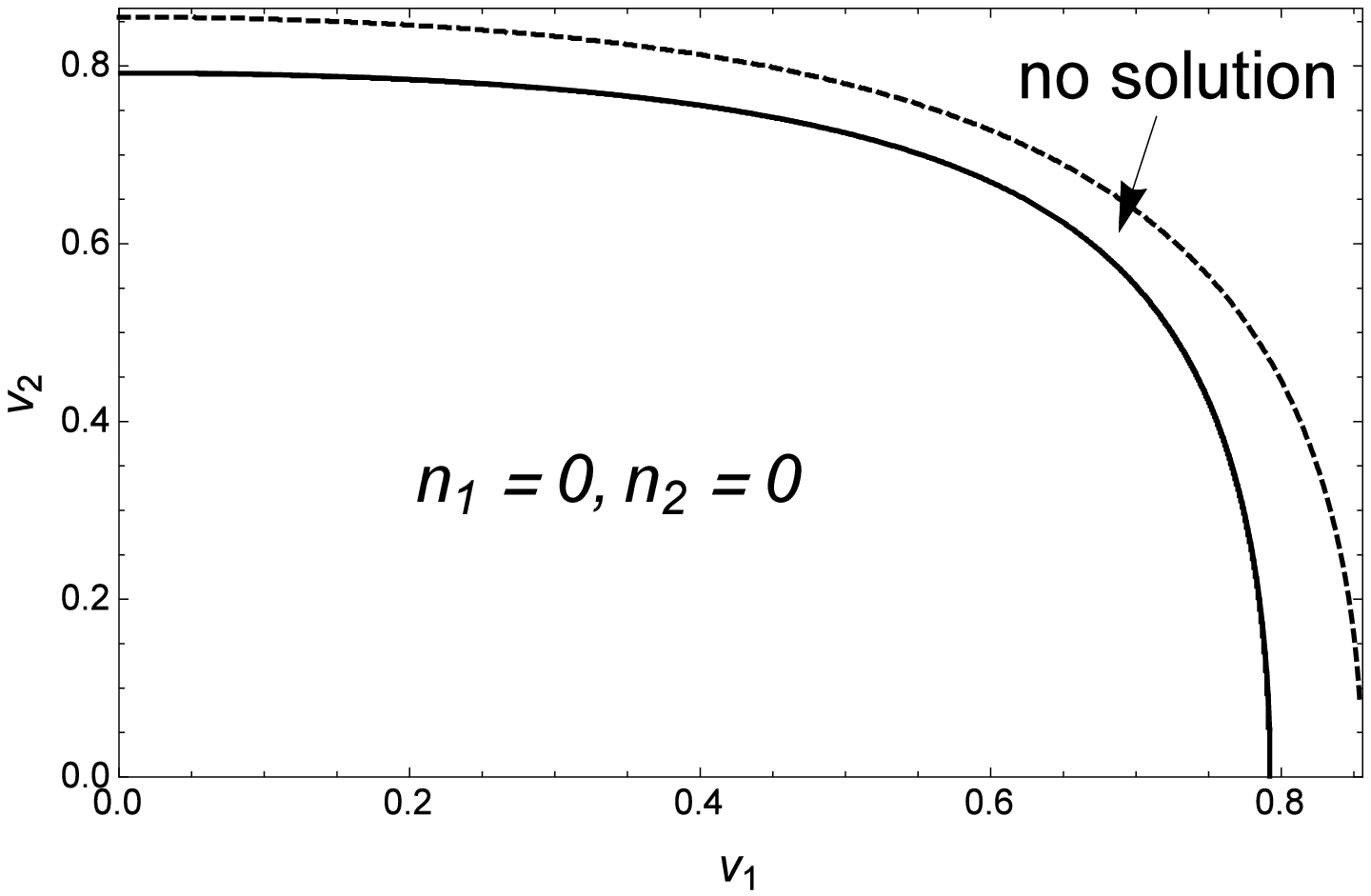}   \\
(a)  $\Lambda = -5.1$ & (b) $\Lambda = -7$ \\
\includegraphics[width=0.95\columnwidth]{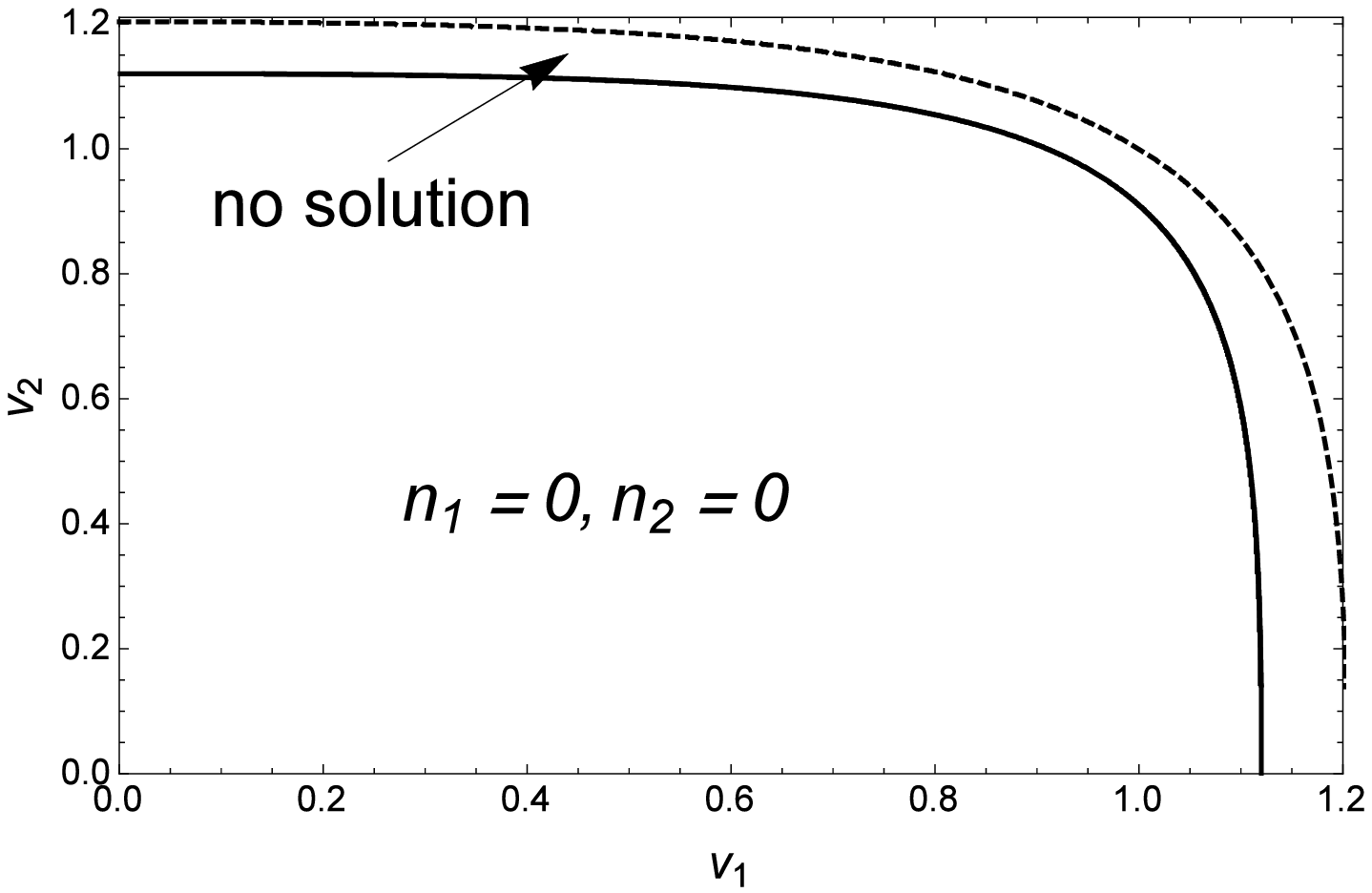} &
\includegraphics[width=0.95\columnwidth]{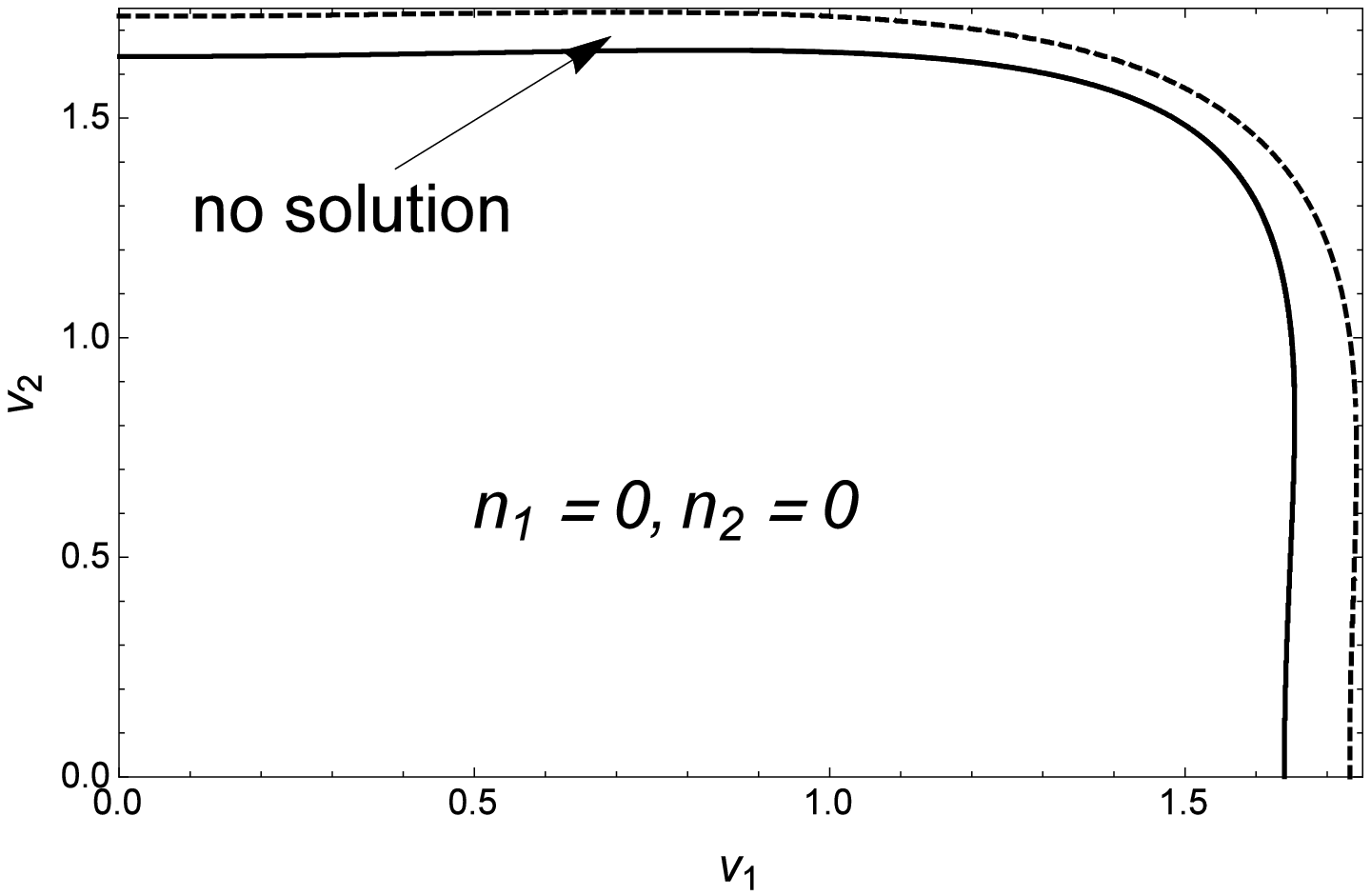}   \\
(c) $\Lambda = -10$ & (d) $\Lambda = -20$
\end{tabular}
\end{center}
\caption{Phase space of topological black hole solutions of ${\mathfrak {su}}(3)$ EYM for $k=-1$. We have fixed the event horizon
radius to be $r_{h}=1$ and  $ \Lambda $ to have the values (a)~$-5.1$, (b)~$-7$, (c)~$-10$ and (d)~$-20$. Black hole solutions exist in the region bounded by the axes and the solid curves. For parameter values outside the solid curves there are no solutions.  The region bounded by dashed and solid curves denotes that region of parameter space where the condition (\ref{eq:EHcond}) for a regular event horizon holds, but we did not find a suitable numerical solution.
Regions where there are nontrivial topological black hole solutions are labelled according to the number of zeros $n_{1}$, $n_{2}$ of the magnetic gauge field functions $\omega _{1}$, $\omega _{2}$ respectively. For all solutions found we have $n_{1}=0=n_{2}$.}
\label{fig:su3k=-1}
\end{figure*}

In figures \ref{fig:su3k=0} and \ref{fig:su3k=-1} we plot the phase space of ${\mathfrak {su}}(3)$ topological black holes for $k=0$ and $k=-1$ respectively.
In each case we fix $r_{h}=1$.
The black holes are now parameterized by $v_{1}=\omega _{1}(r_{h})$, $v_{2}=\omega _{2}(r_{h})$ and $\Lambda $. We fix a few selected values of $\Lambda $ for both $k=0$ and $k=-1$ and plot the phase space in $(v_{1}, v_{2})$ in each case.
As in the ${\mathfrak {su}}(2)$ case, we scan over those values of $(v_{1}, v_{2})$ for which the condition (\ref{eq:EHcond}) is satisfied.

In each plot in figures \ref{fig:su3k=0} and \ref{fig:su3k=-1}, the condition (\ref{eq:EHcond}) for a regular event horizon is satisfied by parameter values in the region enclosed by the axes and the dotted curve.  It can be seen how the size of this region expands as $\left| \Lambda \right| $ increases, in accordance with (\ref{eq:EHcond}).
For some values of the parameters such that (\ref{eq:EHcond}) holds, we are unable to find a suitable numerical solution, and these values of the parameters are in the region labelled ``no solution'' between the dotted and solid curves.  We find nontrivial topological black hole solutions for values of the parameters $(v_{1}, v_{2})$ in the regions bounded by solid curves and the axes.
The regions with solutions are labelled according to the number of zeros $(n_{1}, n_{2})$ of the magnetic gauge field functions $\omega _{1}(r)$, $\omega _{2}(r)$ respectively.

From figures \ref{fig:su3k=0}, \ref{fig:su3k=-1}, the region of parameter space where we have nontrivial topological black hole solutions expands as $\left| \Lambda \right| $ increases, both in absolute size and in its size relative to the size of the region of parameter space for which (\ref{eq:EHcond}) holds.
For both $k=0$ and $k=-1$, embedded ${\mathfrak {su}}(2)$ topological black hole solutions lie on the line $v_{1}=v_{2}$.  From (\ref{eq:su2embedded}) and the discussion in section \ref{sec:su2}, for these embedded solutions the two magnetic gauge field functions are equal and have no zeros so $n_{1}=0=n_{2}$.
Since the field equations (\ref{eq:EE}, \ref{eq:YME}) are invariant under the transformation $j\rightarrow N-j$, the phase space of ${\mathfrak {su}}(3)$ black holes is symmetric about the line $v_{1}=v_{2}$.
The point where $v_{1}=0=v_{2}$ gives the trivial embedded Reissner-Nordstr\"om-adS black hole with magnetic charge $Q=2k$ (\ref{eq:charge}).

For $k=0$, the condition (\ref{eq:EHcond}) makes no constraints on the magnitude of the cosmological constant $\Lambda $ and we find nontrivial topological black holes for all values of $\Lambda $ examined.
For each $\Lambda $ we find a region of nontrivial topological black hole solutions for which both gauge field functions $\omega _{1}$ and $\omega _{2}$ have no zeros, whose existence was proven in \cite{Baxter:2014nka} for sufficiently large $\left| \Lambda \right| $.
This region contains the embedded ${\mathfrak {su}}(2)$ solutions.
For larger values of $\left| \Lambda \right|$, all nontrivial solutions found are such that both $\omega _{1}$ and $\omega _{2}$ have no zeros.

However, for smaller values of $\left| \Lambda \right| $ we find nontrivial topological black hole solutions for which at least one of $\omega _{1}$, $\omega _{2}$ has zeros.
The corresponding regions of parameter space lie between the nodeless $n_{1}=0=n_{2}$ regions (which contain all the embedded ${\mathfrak {su}}(2)$ solutions) and the ``no solution'' region.
When $\Lambda = -1$ we find solutions with $n_{1}=0$, $n_{2}=1$ and $n_{1}=1$, $n_{2}=0$ in very small regions exterior to the $n_{1}=0=n_{2}$ region.
The size of the regions of parameter space containing $n_{1}=1$, $n_{2}=0$ or $n_{1}=0$, $n_{2}=1$ solutions increases as $\left| \Lambda \right| $ decreases, as a proportion of the total size of the region of parameter space for which there are nontrivial solutions.
For $\Lambda = -0.01$ we also find numerical solutions with $n_{1}=2$, $n_{2}=0$ and $n_{1}=0$, $n_{2}=2$, but the relevant regions of parameter space are too small to be seen in figure \ref{fig:su3k=0}(a).  They lie just outside the regions where $n_{1}=1$, $n_{2}=0$ or $n_{1}=0$, $n_{2}=1$.

The overall shape of the region of parameter space for which there are nontrivial solutions with $k=0$ also changes as $\left| \Lambda \right| $ increases: for small $\left| \Lambda \right| $ we see a cusp-like shape at the maximum values of $v_{1}$ and $v_{2}$ in this region, but this smooths out for larger values of $\left| \Lambda \right| $.
This qualitative feature is also seen the $k=1$ spherically symmetric black holes discussed in \cite{Baxter:2007au}.

The phase space shown in figure \ref{fig:su3k=0} is much simpler than that for spherically symmetric solutions with $k=1$ \cite{Baxter:2007au}.
When $k=0$ there are nodeless solutions for any value of $\left| \Lambda \right| $, no matter how small (in a neighbourhood of the embedded ${\mathfrak {su}}(2)$ solutions), whereas when $k=1$ and $\left| \Lambda \right| = 10^{-4}$ there are no nodeless solutions \cite{Baxter:2007au}.
For a fixed $\left| \Lambda \right| $, we also find a smaller range of values of $n_{1}$ and $n_{2}$ when $k=0$ compared with $k=1$.  For example,
with $\left| \Lambda \right| = 0.1$, when $k=0$ we find the following combinations of $(n_{1}, n_{2})$: $(0,0)$, $(1,0)$ and $(0,1)$, but with $k=1$ there are also the combinations $(1,1)$, $(2,0)$, $(2,1)$, $(2,2)$, $(1,2)$ and $(0,2)$ \cite{Baxter:2007au}.
As $\left| \Lambda \right| $ decreases towards zero, in the $k=0$ case presented here we do not see the fragmentation of the phase space seen when $k=1$ \cite{Baxter:2007au}. This is because for $k=0$ there is no solution in the $\left| \Lambda \right| \rightarrow 0 $ limit; instead the phase space simply shrinks in size as $\left| \Lambda \right| $ decreases.

For $k=-1$, with $r_{h}=1$, the minimum value of $\left| \Lambda \right| $ for which there is a regular event horizon is $\left| \Lambda \right| =5$.
Unlike the situation for $k=0$, for $k=-1$ all the topological black hole solutions that we find numerically are such that both magnetic gauge field functions have no zeros, so $n_{1}=0=n_{2}$.
Furthermore, the shape of the region of parameter space for which there are nontrivial solutions does not vary much as $\left| \Lambda \right| $ increases.

\section{Discussion}
\label{sec:conc}

In this letter we have constructed numerical solutions of the ${\mathfrak {su}}(N)$ Einstein-Yang-Mills (EYM) equations in anti-de Sitter (adS) space-time representing purely magnetic topological black holes.  These solutions generalize the numerical spherically-symmetric black holes and solitons discussed in detail in \cite{Baxter:2007au}.  We have explored the phase space of solutions for ${\mathfrak {su}}(2)$ and ${\mathfrak {su}}(3)$ gauge groups, for both $k=0$ (planar event horizon topology) and $k=-1$ (when the event horizon is a surface of constant negative curvature).

In the ${\mathfrak {su}}(2)$ case, the (single) magnetic gauge field function has no zeros for all topological black holes with both $k=0$ and $k=-1$ \cite{VanderBij:2001ia}.
For $N>2$, the existence of topological black holes for which all the magnetic gauge field functions have no zeros is proven in \cite{Baxter:2014nka}.
These solutions are of particular interest because at least some of them have been proven to be stable under linear perturbations of the metric and gauge field \cite{Baxter:2015xfa}.
In the ${\mathfrak {su}}(3)$ case, all the numerical solutions we find for $k=-1$ have nodeless magnetic gauge field functions, but this is not the case for $k=0$.
When $k=0$, as well as the nodeless solutions whose existence was proven in \cite{Baxter:2014nka}, we also find topological black holes for which at least one of the gauge field functions has one or more zeros.

Topological black holes with Ricci-flat ($k=0$) event horizons have attracted a great deal of attention recently as models of holographic superconductors (see, for example, \cite{Cai:2015cya} for a recent review).
In particular, ${\mathfrak {su}}(2)$ EYM black holes in adS with $k=0$ have been studied as gravitational analogues of $p$-wave superconductors \cite{Gubser:2008wv}.
Unlike the situation in this paper (where we consider only purely magnetic gauge fields), in holographic superconductor models the gauge field is dyonic, with nonzero electric and magnetic parts.
The magnetic part of the ${\mathfrak {su}}(2)$ gauge field forms a nontrivial condensate outside the event horizon and vanishes as $r\rightarrow \infty $ \cite{Gubser:2008wv,Gubser:2008zu}.
A natural question is whether enlarging the gauge group to ${\mathfrak {su}}(N)$ yields solutions which also model holographic superconductors.
Recently the existence of dyonic topological black holes in ${\mathfrak {su}}(N)$ EYM in adS was proven \cite{Baxter:2015tda}, but in that paper the solutions shown to exist are such that all the gauge field functions have no zeros (including at infinity), which are not relevant for holographic superconductors.
A numerical study of the solutions of the field equations for a dyonic gauge field, extending the work in this letter, is therefore needed.
We leave an investigation of this for future work.

\section*{Acknowledgments}
The work of E.W.~is supported by the Lancaster-Manchester-Sheffield Consortium for
Fundamental Physics under STFC grant ST/L000520/1.



\begin{thebibliography}{99}

\bibitem{Hawking:1971vc}
  S.W.~Hawking, Commun.~Math.~Phys.~{\bf 25} (1972) 152--166.

\bibitem{Hawking:1973uf}
  S.W.~Hawking and G.F.R.~Ellis,
  The large scale structure of space-time, Cambridge University Press, Cambridge, 1973.

\bibitem{Birmingham:1998nr}
  D.~Birmingham,  Class.~Quant.~Grav.~{\bf 16} (1999) 1197--1205.

\bibitem{Brill:1997mf}
  D.R.~Brill, J.~Louko and P.~Peldan,  Phys.~Rev.~D {\bf 56} (1997) 3600--3610.

\bibitem{Lemos:1994fn}
  J.P.S.~Lemos,  Class.~Quant.~Grav.~{\bf 12} (1995) 1081--1086.

\bibitem{Lemos:1994xp}
  J.P.S.~Lemos,  Phys.~Lett.~B {\bf 353} (1995) 46--51.

\bibitem{Lemos:1995cm}
  J.P.S.~Lemos and V.T.~Zanchin, Phys.~Rev.~D {\bf 54} (1996) 3840--3853.

\bibitem{Vanzo:1997gw}
  L.~Vanzo,  Phys.~Rev.~D {\bf 56} (1997) 6475--6483.

\bibitem{Cai:1996eg}
  R.G.~Cai and Y.Z.~Zhang,
    Phys.~Rev.~D {\bf 54} (1996) 4891--4898.

\bibitem{Mann:1996gj}
  R.B.~Mann,  Class.~Quant.~Grav.~{\bf 14} (1997) L109--L114.

\bibitem{Smith:1997wx}
  W.L.~Smith and R.B.~Mann,
  Phys.~Rev.~D {\bf 56} (1997) 4942--4947.

\bibitem{Mann:1997zn}
  R.B.~Mann,  Nucl.~Phys.~B {\bf 516} (1998) 357--381.

\bibitem{Bizon:1990sr}
  P.~Bizon,  Phys.~Rev.~Lett.~{\bf 64} (1990) 2844--2847.

\bibitem{Straumann:1990as}
  N.~Straumann and Z.~H.~Zhou,  Phys.~Lett.~B {\bf 243} (1990) 33--35.

\bibitem{Winstanley:1998sn}
  E.~Winstanley,  Class.~Quant.~Grav.~{\bf 16} (1999) 1963--1978.

\bibitem{Bjoraker:1999yd}
  J.~Bjoraker, Y.~Hosotani,  Phys.~Rev.~Lett.~{\bf 84} (2000) 1853--1856.

\bibitem{Bjoraker:2000qd}
  J.~Bjoraker, Y.~Hosotani, Phys.~Rev.~D {\bf 62} (2000) 043513.

\bibitem{VanderBij:2001ia}
  J.J.~Van der Bij, E.~Radu,  Phys.~Lett.~B {\bf 536} (2002) 107--113.

\bibitem{Winstanley:2008ac}
  E.~Winstanley,  Lect.~Notes Phys.~{\bf 769} (2009) 49--87.

\bibitem{Winstanley:2015loa}
  E.~Winstanley,  {\tt {arXiv:1510.01669 [gr-qc]}}.

\bibitem{Baxter:2007at}
  J.E.~Baxter, M.~Helbling, E.~Winstanley, Phys.~Rev.~Lett.~{\bf 100} (2008) 011301.

\bibitem{Baxter:2008pi}
  J.E.~Baxter, E.~Winstanley, Class.~Quant.~Grav.~{\bf 25} (2008) 245014.

\bibitem{Baxter:2015gfa}
  J.E.~Baxter, E.~Winstanley, {\tt {arXiv:1501.07541 [gr-qc]}}.

\bibitem{Baxter:2014nka}
  J.E.~Baxter, Gen.~Rel.~Grav.~{\bf 47} (2015) 1829.

\bibitem{Baxter:2015xfa}
  J.E.~Baxter, {\tt {arXiv:1507.03127 [gr-qc]}}.

\bibitem{Baxter:2007au}
  J.E.~Baxter, M.~Helbling, E.~Winstanley, Phys.~Rev.~D {\bf 76} (2007) 104017.

\bibitem{Kunzle:1991}
  H.P.~Kunzle,  Class.~Quant.~Grav.~{\bf {8}} (1991) 2283–-2297.

\bibitem{Cai:2015cya}
  R.G.~Cai, L.~Li, L.F.~Li, R.Q.~Yang,  Sci.~China Phys.~Mech.~Astron.~{\bf 58} (2015) 060401.

\bibitem{Gubser:2008wv}
  S.S.~Gubser and S.S.~Pufu,  JHEP {\bf 0811} (2008) 033.

\bibitem{Gubser:2008zu}
  S.S.~Gubser, Phys.~Rev.~Lett.~{\bf 101} (2008) 191601.

\bibitem{Baxter:2015tda}
  J.E.~Baxter, {\tt {arXiv:1507.05314 [gr-qc]}}.


\end{thebibliography}
\end{document}